\documentclass{article}

\pdfoutput=1
\usepackage{PRIMEarxiv}
\usepackage{cite}
\usepackage[utf8]{inputenc} 
\usepackage[T1]{fontenc}    
\usepackage{hyperref}       
\usepackage{url}            
\usepackage{booktabs}       
\usepackage{amsfonts}       
\usepackage{nicefrac}       
\usepackage{microtype}      
\usepackage{lipsum}
\usepackage{fancyhdr}       
\usepackage{graphicx}       
\usepackage{amsmath}
\graphicspath{{media/}}     

\title{PF-DMD: Physics-fusion dynamic mode decomposition for accurate and robust forecasting of dynamical systems with imperfect data and physics
}

\author{
  Yuhui Yin\thanks{Yuhui Yin and Chenhui Kou contributed equally to the article.} \\
  State Key Laboratory of Chemical Engineering \\
  Tianjin University \\
  Tianjin\\
  \texttt{yinyuhui@tju.edu.cn} \\
   \And
  Chenhui Kou\footnotemark[1] \\
  State Key Laboratory of Chemical Engineering, Department of Statistics and Data Science \\
  Tianjin University, Yale University \\
  Tianjin, New Haven\\
  \texttt{chenhui.kou@yale.edu} \\
   \And
  Shengkun Jia\footnotemark[2] \\
  State Key Laboratory of Chemical Engineering \\
  Tianjin University \\
  Tianjin\\
  \texttt{jiask@tju.edu.cn} \\
   \And
  Lu Lu \\
  Department of Statistics and Data Science \\
  Yale University \\
  New Haven\\
  \texttt{lu.lu@yale.edu} \\
   \And
  Xigang Yuan\footnotemark[2] \\
  State Key Laboratory of Chemical Engineering \\
  Tianjin University \\
  Tianjin\\
  \texttt{yuanxg@tju.edu.cn} \\
   \And
  Yiqing Luo \\
  State Key Laboratory of Chemical Engineering \\
  Tianjin University \\
  Tianjin\\
  \texttt{luoyq@tju.edu.cn} \\
}

\begin{document}
\maketitle

\begin{abstract}
The DMD (Dynamic Mode Decomposition) method has attracted widespread attention as a representative modal-decomposition method and can build a predictive model. However, the DMD may give predicted results that deviate from physical reality in some scenarios, such as dealing with translation problems or noisy data. Therefore, this paper proposes a physics-fusion dynamic mode decomposition (PFDMD) method to address this issue. The proposed PFDMD method first obtains a data-driven model using DMD, then calculates the residual of the physical equations, and finally corrects the predicted results using Kalman filtering and gain coefficients. In this way, the PFDMD method can integrate the physics-informed equations with the data-driven model generated by DMD. Numerical experiments are conducted using the PFDMD, including the Allen-Cahn, advection-diffusion, and Burgers' equations. The results demonstrate that the proposed PFDMD method can significantly reduce the reconstruction and prediction errors by incorporating physics-informed equations, making it usable for translation and shock problems where the standard DMD method has failed. 
\end{abstract}

\keywords{Dynamic mode decomposition \and Data-driven \and Partial differential equation}

\section{Introduction}
In recent years, modal-decomposition methods have rapidly developed and succeeded in fluid dynamics\cite{r1,r2,r3}. At first, modal decomposition is used as a mathematical tool to analyze data for fluid dynamics. However, many researchers have conceived that it can also be used in system identification - a process by which a model is constructed for a system from measurement data, and some can make great predictions. It enables fast and efficient data reduction\cite{r4,r5}, data analysis\cite{r6,r7,r8}, and reduced-order modeling (ROM) \cite{r9,r10,r11} of complex datasets. Besides effectively analyzing experimental and simulation data, model decomposition methods can also predict fluid dynamics by extracting underlying physical patterns. Among them, Proper Orthogonal Decomposition (POD) \cite{r12,r13} and Dynamic Mode Decomposition (DMD) \cite{r4,r6,r14} are widely used to construct data-driven models in scientific and engineering fields. 

The POD was brought to the fluid dynamics first by Lumley J \cite{r15}. It is a data decomposition from a dynamical system into a hierarchical set of orthogonal modes according to their energy content. However, POD is an energy-based modal-decomposition method and cannot separate spatial and temporal signals, so additional processing is required for data-driven prediction \cite{r16}. DMD is an alternative method based on the Koopman operator \cite{r6}. The infinite-dimensional Koopman mode can be truncated and approximated using the DMD algorithm proposed by Schmid \cite{r4}. The main advantage of DMD is it can decompose high-dimensional spatiotemporal signals into a triplet of purely spatial modes, scalar amplitudes, and purely temporal signals in a process. The foundation of this decomposition allows us to construct and predict the full data sequence. Therefore, DMD can extract low-dimensional non-orthogonal modes from simulation and experiment time-series data snapshots. These extracted modes can be used for model construction and analysis of continuous unsteady nonlinear flow fields. So far, POD and DMD methods have been applied in various areas. Freund \cite{r17} used different norms to reveal the flow structures associated with the sound field. Raiola \cite{r18} proposed a filtering method based on POD to improve the physical insight into the investigation of turbulent flows. Han \cite{r19} captured the coherent structures of the tip leakage vortex in a mixed-flow pump using DMD and reconstructed the corresponding flow field. Kanbur \cite{r20} successfully predicted the surface temperature and heat discharge rate of lithium polymer batteries in electrochemistry using DMD. Zhao \cite{r21} extracted the energy and dynamic characteristics of the high-shear mixer velocity field in a high-shear force mixer problem in fluid mechanics using POD and DMD. Amor \cite{r22} utilized a higher order DMD to identify the main patterns describing the flow motion in complex flows and enhanced it by implementing a sliding window to speed up calculation. 

As a data-driven prediction tool, the DMD method still has some limitations in practical applications. Firstly, DMD fundamentally relies on the variable separation of Singular Value Decomposition (SVD) \cite{r4}, which requires that the observed data must have a certain degree of alignment \cite{r23}. This means that measurement points in an experiment or simulation should remain aligned over time, without rotating or translating. Otherwise, DMD cannot capture appropriate low-dimensional modes that satisfy physical constraints. This limitation makes it difficult to apply the DMD method to predict translation systems such as advection-dominated diffusion \cite{r24}and bubble dynamics \cite{r10}. Secondly, DMD is sensitive to noise in data \cite{r25}. Errors caused by noise accumulate during the prediction, eventually making the constructed model lose its predictive ability. Lastly, it is worth noting that DMD, as a purely data-driven method, lacks any priori assumptions or knowledge of the underlying dynamics. This can result in poor generalization of the obtained model, as it lacks physical support. In fluid systems with strong nonlinearity and coupling, this limitation often leads to the failure of the DMD method.

In order to address these limitations, researchers have made some improvements to the DMD method. To overcome the limitation of requiring alignment of DMD and the failure of the translation problem, Lu \cite{r24} integrated the Lagrangian framework into DMD and supplemented the velocity data to the dataset, successfully predicting the translation problem. Using Lagrangian DMD, Yin \cite{r10} accurately reconstructed and predicted the two-dimensional bubble-rising process. Regarding the error caused by noise, Dawson \cite{r26} and Hemati \cite{r27} argued that DMD has difficulties extracting accurate dynamic features from noise-corrupted data and proposed various modification strategies. Lastly, to introduce prior physical model features, Baddoo \cite{r28} utilized the matrix manifold corresponding to physical laws as prior knowledge, purposefully decomposing the linear coefficient matrix into manifolds to enhance the application capabilities of DMD. Gao \cite{r29} proposed a DMD variant based on the k-nearest-neighbors (KNN) to the applicability of DMD to parameterized problems. To cope with inhomogeneous problems, Lu \cite{r30} introduced a bias term and identity mapping into DMD. However, these methods are only effective for specific forms of physical information. Therefore, there is a need for a more generalized method to integrate prior knowledge, such as physical equations, into the DMD method to address violations of physical constraints, as well as improve model generalization and robustness.

In this paper, we propose a Physics-Fusion Dynamic Mode Decomposition (PFDMD) method that can conveniently integrate data-driven models and physical equations. Based on the application of data assimilation, The proposed PFDMD method introduces more general physical equations through Kalman filtering to correct the linear models generated solely from data-driven approaches. The proposed PFDMD method is elaborated in Section 2. In Section 3, the PFDMD method is tested on well-known equations like the advection-diffusion, Allen–Cahn, and Burgers’ equations, some of which are typically hard to predict with the standard DMD method. The predictive accuracy of the PFDMD method is evaluated, and a balance between physics and data model is demonstrated. We also assess the robustness of PFDMD when dealing with noisy data, inaccurate physical information, and limited data. In Section 4, conclusions and prospects are provided.

\section{Methods}
\subsection{Dynamic Mode Decomposition}

This section provides a brief introduction to the DMD method in dynamic system. In a dynamic system, the observation data (measurement points in experiments or grid points in numerical simulations) at a specific moment ${t}_{k}$ are referred to as snapshots and denoted as ${x}_{k}$. By reshaping the snapshot data into a column vector, the current ${x}_{k}$ and next ${x}_{k+1}$ snapshots are paired to form a time series: $\left\{ \left( {{x}_{k}},{{x}_{k+1}} \right) \right\}_{k=1}^{m},{{t}_{k+1}}={{t}_{k}}+\Delta t$. In the DMD method, these snapshots are arranged according to the time series and form two matrices ${{X}_{1}}$ and ${{X}_{2}}$:
\begin{equation}\label{eq1}
{{X}_{1}}=\left[ \begin{matrix}
   \begin{matrix}
{{x}_{1}} & {{x}_{2}} & \cdots  & {{x}_{n-1}}  \\
\end{matrix}  \\
\end{matrix} \right]\in {{\mathbb{R}}^{m\times \left( n-1 \right)}}
\end{equation}
\begin{equation}\label{eq2}
{{X}_{2}}=\left[ \begin{matrix}
   \begin{matrix}
   {{x}_{2}} & {{x}_{3}} & \cdots  & {{x}_{n}}  \\
\end{matrix}  \\
\end{matrix} \right]\in {{\mathbb{R}}^{m\times \left( n-1 \right)}}
\end{equation}

The DMD method is based on the linear assumption and aims to find the best-fit linear operator between two matrices:
\begin{equation}\label{eq3}
{{X}_{2}}=A{{X}_{1}}
\end{equation}
\begin{equation}\label{eq4}
A=\underset{A}{\mathop{\arg \min }}\,{{\left\| {{X}_{2}}-A{{X}_{1}} \right\|}_{F}}={{X}_{2}}X_{1}^{\dagger }
\end{equation}
where ${{\left\| \,\,\cdot \,\, \right\|}_{F}}$ represents the Frobenius norm, and $\dagger$ denotes the pseudo-inverse. In computations, due to the high dimensionality of the data, computing the inverse directly using the least squares method requires significant computational resources. Therefore, we obtain the pseudo-inverse through SVD decomposition:
\begin{equation}\label{eq5}
{{X}_{1}}\approx U\Sigma {{V}^{*}}
\end{equation}
where $\ast$ denotes the conjugate transpose. In addition to computing the pseudo-inverse, SVD can also approximate the rank of the data matrix \emph{X}, thereby reducing dimensionality and computational complexity. After performing SVD, the approximate matrix of \emph{X} can be obtained as $\tilde{A}$, which is the result of data dimensionality reduction:
\begin{equation}\label{eq6}
A={{X}_{2}}V{{\Sigma }^{-1}}{{U}^{*}}
\end{equation}
\begin{equation}\label{eq7}
\tilde{A}={{U}^{*}}AU={{U}^{*}}{{X}_{2}}V{{\Sigma }^{-1}}
\end{equation}
Afterward, spectral decomposition is performed on the $\tilde{A}$ matrix, obtaining the eigenvalues and the eigenvectors:
\begin{equation}\label{eq8}
\tilde{A}W=W\Lambda
\end{equation}
where the column vectors of \emph{W} are the eigenvectors of $\tilde{A}$, and the diagonal $\Lambda$ consists of the corresponding eigenvalues $\lambda$.

Then the DMD modes is introduced based on the DMD method, which are reconstructed by combining the next timestep ${{X}_{2}}$ matrix with the eigenvalue matrix \emph{W} and constitute the mode decomposition of the \emph{A} matrix.
\begin{equation}\label{eq9}
\Phi ={{X}_{2}}V{{\Sigma }^{-1}}W
\end{equation}
\begin{equation}\label{eq10}
A\Phi =\left( {{X}_{2}}V{{\Sigma }^{-1}}{{U}^{*}} \right)\left( {{X}_{2}}V{{\Sigma }^{-1}}W \right)=\Phi \Lambda 
\end{equation}
Once $\Lambda$ is obtained, it allows for the reconstruction and prediction of any snapshot at any given time:
\begin{equation}\label{eq11}
{{x}_{k}}=\Phi {{\Lambda }^{k-1}}{{\Phi }^{\dagger }}{{x}_{1}}
\end{equation}

\subsection{Physics-Fusion DMD method}
\begin{figure}
  \centering
  \includegraphics[width=\textwidth]{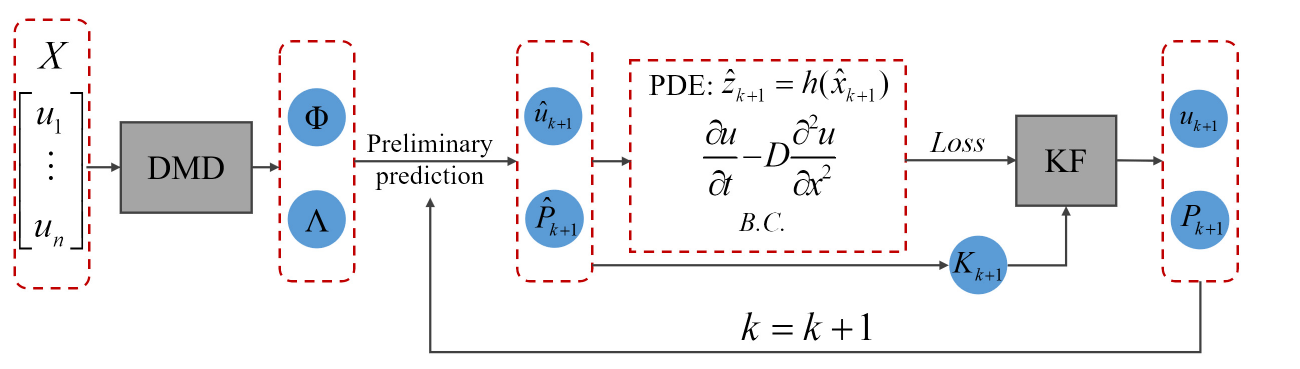}
  \caption{The algorithm flowchart of PFDMD.}
  \label{fig:fig1}
\end{figure}

DMD method can construct a linear model through data, with the key step being the derivation of the linear mapping matrix A and its mode decomposition. Based on this decomposition result, it is believed that a preliminary linear approximation model of the system has been obtained. However, as mentioned in the introduction, this purely data-driven model often fails to capture dynamic mapping relationship due to the lack of prior knowledge and physical information, thus constructing a model away from physical reality. Therefore, to address this deficiency, this paper proposes a Physics-Fusion Dynamic Mode Decomposition (PFDMD) method, which integrates physical information and a data-driven model. The algorithm flowchart of PFDMD is shown in Fig. \ref{fig:fig1}.

As shown in Fig.\ref{fig:fig1}, after obtaining the \emph{A} matrix and decomposing to $\Phi$ and $\Lambda$ through DMD, we can introduce physical information into the model by using the Kalman Filter framework to correct the predicted results with the residual of the physical equations. First, we can preliminarily predict the state at time \emph{k}+1:
\begin{equation}\label{eq12}
{{\hat{u}}_{k+1}}=A{{u}_{k}}=\Phi \Lambda {{\Phi }^{\dagger }}{{u}_{k}}
\end{equation}
where ${{u}_{k}}$ represents the snapshot at the current time, and ${{\hat{u}}_{k+1}}$ represents the preliminarily predicted snapshot for the next time step obtained through DMD. After obtaining the preliminary prediction result, we bring them into the physical equations and correct the prediction result by the equation residuals. Therefore, the predicted residuals of the physical information are expressed as follows: 
\begin{equation}\label{eq13}
{{\hat{z}}_{k+1}}=h({{\hat{u}}_{k+1}})
\end{equation}
where $h\left( \cdot  \right)$ represents the physical information operator, the physical equations the snapshot should satisfy. Afterward, as shown in Fig.1, the residual of the physical information $\left( {{z}_{k+1}}-{{{\hat{z}}}_{k+1}} \right)$ is corrected using a Kalman gain coefficient \emph{K} and then applied to the predicted results. 
\begin{equation}\label{eq14}
{{u}_{k+1}}={{\hat{u}}_{k+1}}+{{K}_{k+1}}\left( {{z}_{k+1}}-{{{\hat{z}}}_{k+1}} \right)={{\hat{u}}_{k+1}}-{{K}_{k+1}}{{\hat{z}}_{k+1}}
\end{equation}
where ${{u}_{k+1}}$ is the predicted result after the correction of physical information, \emph{K} is the Kalman gain coefficient, ${{z}_{k+1}}$ is the theoretical residual of the physical equation, whose value is typically zero. ${{\hat{z}}_{k+1}}$ is the predicted residual of the physical equations. So, we can get the simplified expression on the right-hand side of Eq. (\ref{eq14}).

The Kalman gain can be obtained through the covariance matrix P. The preliminary prediction of the covariance matrix P is expressed as follows:
\begin{equation}\label{eq15}
{{\hat{P}}_{k+1}}=\Phi \Lambda {{\Phi }^{\dagger }}{{P}_{k}}{{\left( \Phi \Lambda {{\Phi }^{\dagger }} \right)}^{T}}+Q
\end{equation}
where ${{P}_{k}}$ is the snapshot covariance matrix at the current time step, ${{\hat{P}}_{k+1}}$ is the preliminarily predicted covariance matrix at the next step. \emph{Q} represents the process error matrix in the Kalman filtering algorithm, while in this article, it denotes the error of the data-driven model. After predicting the covariance matrix, the Kalman gain can be calculated: 
\begin{equation}\label{eq16}
{{K}_{k+1}}={{\hat{P}}_{k+1}}{{H}_{k+1}}^{T}{{\left[ {{H}_{k+1}}{{{\hat{P}}}_{k+1}}{{H}_{k+1}}^{T}+R \right]}^{-1}}
\end{equation}
\begin{equation}\label{eq17}
{{H}_{k+1}}=\frac{\partial h}{\partial u}=\left[ \begin{matrix}
   \frac{\partial {{h}_{1}}}{\partial {{u}_{1}}} & \frac{\partial {{h}_{1}}}{\partial {{u}_{2}}} & \cdots  & \frac{\partial {{h}_{1}}}{\partial {{u}_{n}}}  \\
   \frac{\partial {{h}_{2}}}{\partial {{u}_{1}}} & \frac{\partial {{h}_{2}}}{\partial {{u}_{2}}} & \cdots  & \frac{\partial {{h}_{2}}}{\partial {{u}_{n}}}  \\
   \vdots  & \vdots  & \vdots  & \vdots   \\
   \frac{\partial {{h}_{n}}}{\partial {{u}_{1}}} & \frac{\partial {{h}_{n}}}{\partial {{u}_{2}}} & \cdots  & \frac{\partial {{h}_{n}}}{\partial {{u}_{n}}}  \\
\end{matrix} \right]
\end{equation}
where \emph{K} represents the Kalman gain coefficient, and \emph{R} represents the measurement error matrix in the Kalman filter, which represents the error of the physical information in this article. \emph{H} is the Jacobian matrix of the nonlinear function $h\left( \cdot  \right)$. By Eq. (\ref{eq14}), the predicted result can be corrected by combining the physical information. At this point, the PFDMD completes the correction of the predicted result for the next time step and then updates the \emph{P} matrix for the next iteration:
\begin{equation}\label{eq18}
{{P}_{k+1}}=\left[ I-{{K}_{k+1}}{{H}_{k+1}} \right]{{\hat{P}}_{k+1}}
\end{equation}

By iteration, all the predicted results can be obtained after the correction by physical information. It is worth noting that the prediction process does not introduce new measurement data. Finally, in Eqs. (\ref{eq15}) and (\ref{eq16}), there are two parameters, \emph{Q} and \emph{R}, representing the errors of data and physical information, respectively. This article combines \emph{Q} and \emph{R} into a weight coefficient \emph{w}, which measures whether the PFDMD algorithm is closer to the data-driven model or the physical information. The value range is from 0 to 1.
\begin{equation}\label{eq19}
w=\frac{R}{R+Q}.
\end{equation}
When the weight coefficient is small, the error of the physical information R has a smaller proportion, and the predicted results of PFDMD are closer to the physical information. Conversely, a larger weight coefficient indicates that the data-driven model is more significant, and the predicted results are closer to the results of traditional DMD. In this way, by adjusting the value of the weight coefficient, we can control whether the results of PFDMD are closer to the physical information or the data.

\section{Results}
In this section, the proposed PFDMD method is applied to solve partial differential equations (PDEs) to demonstrate its effectiveness and characteristics. In Section 3.1, the PFDMD method is applied to a one-dimensional diffusion equation and an Allen-Cahn equation to show the effects of prediction and reconstruction. The advection-diffusion equation and Burgers’ equation are used as examples in Section 3.2 to illustrate the performance of PFDMD to solving translation and shock wave problems, where standard DMD fails. In section 3.3, the PFDMD method is applied to several special application scenarios: noisy data, inaccurate physical information and an absence of data model. Section 3.4 discussed the update of Kalman gain in certain steps and the computational cost of the algorithm. All examples are compared with the numerical solutions of the PDE equations, and accuracy is measured by the L2 relative error.
\begin{equation}\label{eq20}
Error=\frac{\sqrt{\sum\limits_{i=1}^{N}{{{\left| \bar{u}\left( {{x}_{i}},{{t}_{i}} \right)-u\left( {{x}_{i}},{{t}_{i}} \right) \right|}^{2}}}}}{\sqrt{\sum\limits_{i=1}^{N}{{{\left| u\left( {{x}_{i}},{{t}_{i}} \right) \right|}^{2}}}}}.
\end{equation}
\subsection{Simple diffusion problem}
First, the effectiveness of the PFDMD method is demonstrated through the following simple one-dimensional diffusion equation:
\begin{equation}\label{eq21}
\begin{split}
  & \frac{\partial u}{\partial t}=D\frac{{{\partial }^{2}}u}{\partial {{x}^{2}}},\quad x\in [0,2],\quad t\in [0,2] \\ 
 & b.c.\quad u\left( 0,t \right)=u\left( 2,t \right)=0, \\ 
 & i.c.\quad u\left( x,0 \right)=0.5\exp [-{{\left( x-1 \right)}^{2}}/{{0.05}^{2}}], \\ 
\end{split}
\end{equation}
In this equation, the diffusion coefficient \emph{D} is set to 0.01. The spatial domain is discretized into \emph{m} = 200 intervals, while the time domain is discretized into \emph{n} = 500 steps. In this paper, PDEs are solved with a first-order finite-difference upwind scheme for the advection part and finite center difference for the diffusion part. Both DMD and PFDMD algorithms select the first \emph{n} = 200 snapshots as the dataset to obtain the matrix \emph{A} of the linear equation using Eq. (\ref{eq7}). Therefore, in this study, the results from 0 to 200 snapshots are referred to as "reconstruction", while the results from 200 to 500 snapshots are referred to as "prediction". To ensure a fair comparison, both DMD and PFDMD methods in this study use the same SVD truncation. In this case, SVD truncation is 5 (rank = 5) and already has a great reconstruction and prediction ability.
\begin{figure}
  \centering
  \includegraphics[width=\textwidth]{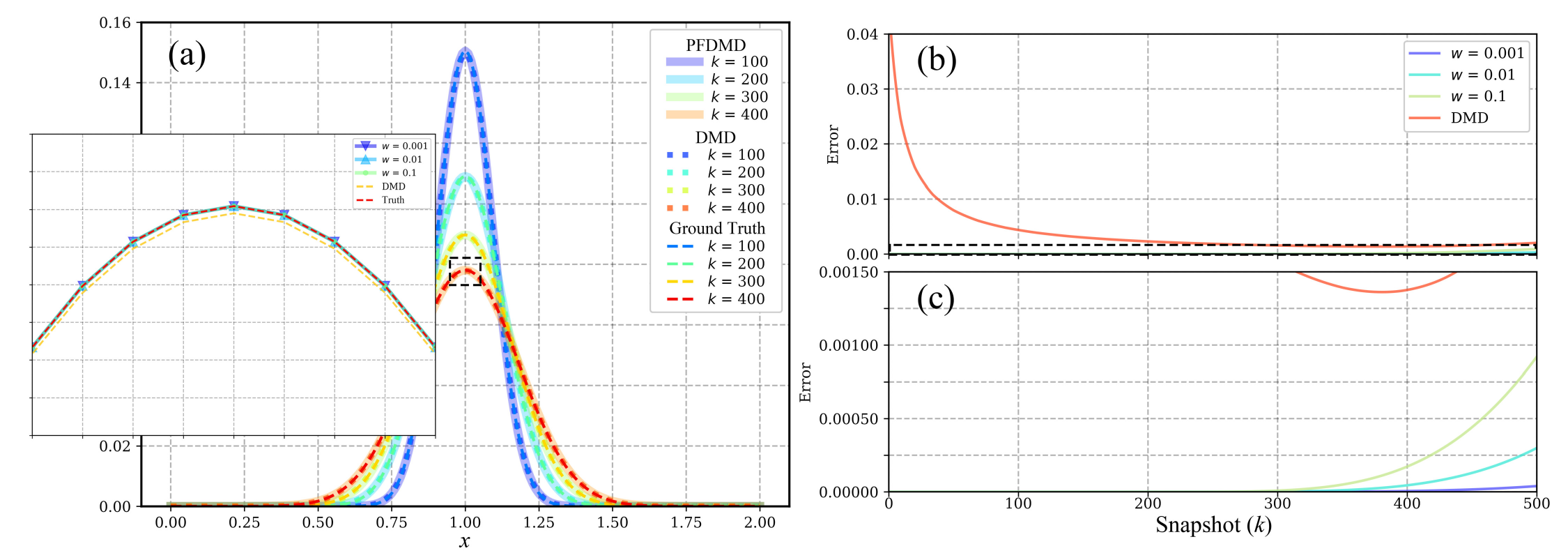}
  \caption{Diffusion equation results: (a) Prediction comparison plot of DMD and PFDMD; (b) Model error plot with different weights; (c) Zoomed-in error plot. (\emph{k} represent the specific snapshot number).}
  \label{fig:fig2}
\end{figure}

The results of the one-dimensional diffusion equation are shown in Fig. \ref{fig:fig2}(a), consistent with the results in Lu's research\cite{r24}. The DMD method has demonstrated excellent predictive capability. However, the zoomed-in detail figure shows that incorporating the physical information in PFDMD results in predictions closer to true values than DMD. Fig. \ref{fig:fig2}(b) demonstrates the reconstruction and prediction errors of the two methods, revealing that the relative error in reconstruction is significantly lower than that in DMD as long as physical information is included. This demonstrates that after integrating the physical information, a better reconstruction result can be obtained by adjusting the coefficients of the DMD-generated model through Kalman filtering. The locally zoomed-in Fig. \ref{fig:fig2}(c) shows that both methods can achieve satisfactory prediction results after 200 snapshots. Still, the prediction errors increase over time due to the error accumulation caused by the linear model. Furthermore, from the results in Fig. \ref{fig:fig2}(c), it can be observed that the smaller the weight coefficient w, the greater the physical information impact and the better the predictive capability of the model.

\begin{figure}
  \centering
  \includegraphics[width=\textwidth]{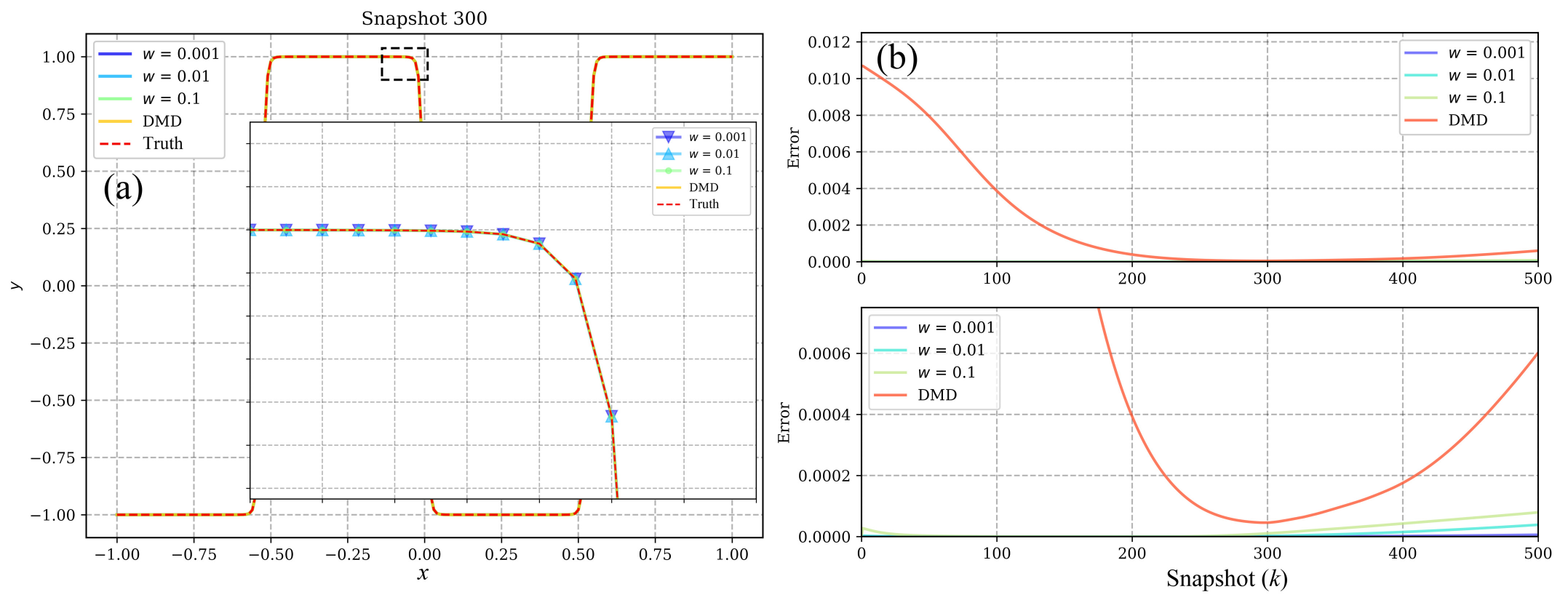}
  \caption{Allen-Cahn equation results: (a) Ground truth and prediction of snapshot 300; (b) Model error plot with different weights.}
  \label{fig:fig3}
\end{figure}

In an alternative case, we take the Allen-Cahn equation in reaction-diffusion as an example to investigate the application of the proposed PFDMD method in more complex nonlinear partial differential equations. Here, we consider the equation with Neumann boundary conditions, which is given as follows:
\begin{equation}\label{eq22}
\begin{split}
  & \frac{\partial u}{\partial t}=\mu \frac{{{\partial }^{2}}u}{\partial {{x}^{2}}}+5\left( u-{{u}^{3}} \right),\quad x\in [-1,1],\quad t\in [0,2] \\ 
 & b.c.\quad {{\left. \frac{\partial u}{\partial x} \right|}_{x=-1}}={{\left. \frac{\partial u}{\partial x} \right|}_{x=1}}=0, \\ 
 & i.c.\quad u\left( x,0 \right)=0.53x+0.47\sin \left( -\frac{3}{2}\pi x \right), \\     
\end{split}
\end{equation}
In this case, the spatial domain [-1,1] is discretized into \emph{m} = 200 intervals, and the time domain [0,2] is discretized into \emph{n} = 500 steps. Similarly, \emph{n} = 200 snapshots are selected as the training set, and the prediction is made up to the $500^{th}$ step. The SVD truncation is 10 (rank = 10). In this example, we set the diffusion coefficient to $\mu =1\times {{10}^{-4}}$. The computed results and errors are shown in Fig. \ref{fig:fig3}. PFDMD exhibits similar performance to one-dimensional diffusion equation, demonstrating that this method can also be applied to more complex semi-linear PDE. 

\subsection{Translation and shock wave problem}
As stated in the introduction, DMD struggles to accurately generate reduced-order models (ROMs) for advection-dominated problems due to its SVD-based approach. The main reason is that the advection-dominated motion involves translation relationships between the snapshots, while the SVD method requires spatial alignment, making it challenging to capture the underlying dynamics throughout the high-dimensional space \cite{r23}. However, in PFDMD, introducing covariance allows for incorporating the relationships between spatial positions. Consequently, PFDMD holds promise in addressing the issue of DMD in advection-dominated processes. In this section, we demonstrate the reconstructive and predictive capability of PFDMD for the advection-diffusion equation and the Burgers’ equation and its two-dimensional scenario.

By adding an advection term to the diffusion equation, we can obtain an advection-dominated problem, translating the \emph{u} distribution over time. It can be described by the following equation:
\begin{equation}\label{eq23}
\begin{split}
  & \frac{\partial u}{\partial t}+a\frac{\partial u}{\partial x}=D\frac{{{\partial }^{2}}u}{\partial {{x}^{2}}},\quad x\in [0,2],\quad t\in [0,2] \\ 
 & b.c.\quad u\left( 0,t \right)=u\left( 2,t \right)=0, \\ 
 & i.c.\quad u\left( x,0 \right)=0.5\exp [-{{\left( x-1 \right)}^{2}}/{{0.05}^{2}}], \\ 
\end{split}
\end{equation}
In this case, the spatial domain [0,2] is discretized into \emph{m} = 200 intervals, and the time domain [0,2] is discretized into \emph{n} = 500 steps. Similarly, \emph{n} = 200 snapshots are selected as the training set, and the prediction is made up to the $500^{th}$ step. In this example, we set the advection and diffusion coefficients as \emph{a} = 0.6, \emph{D} = 0.01, respectively, to increase the proportion of the advection term in the advection-diffusion equation and highlight its translation wave characteristic.
\begin{figure}
  \centering
  \includegraphics[width=\textwidth]{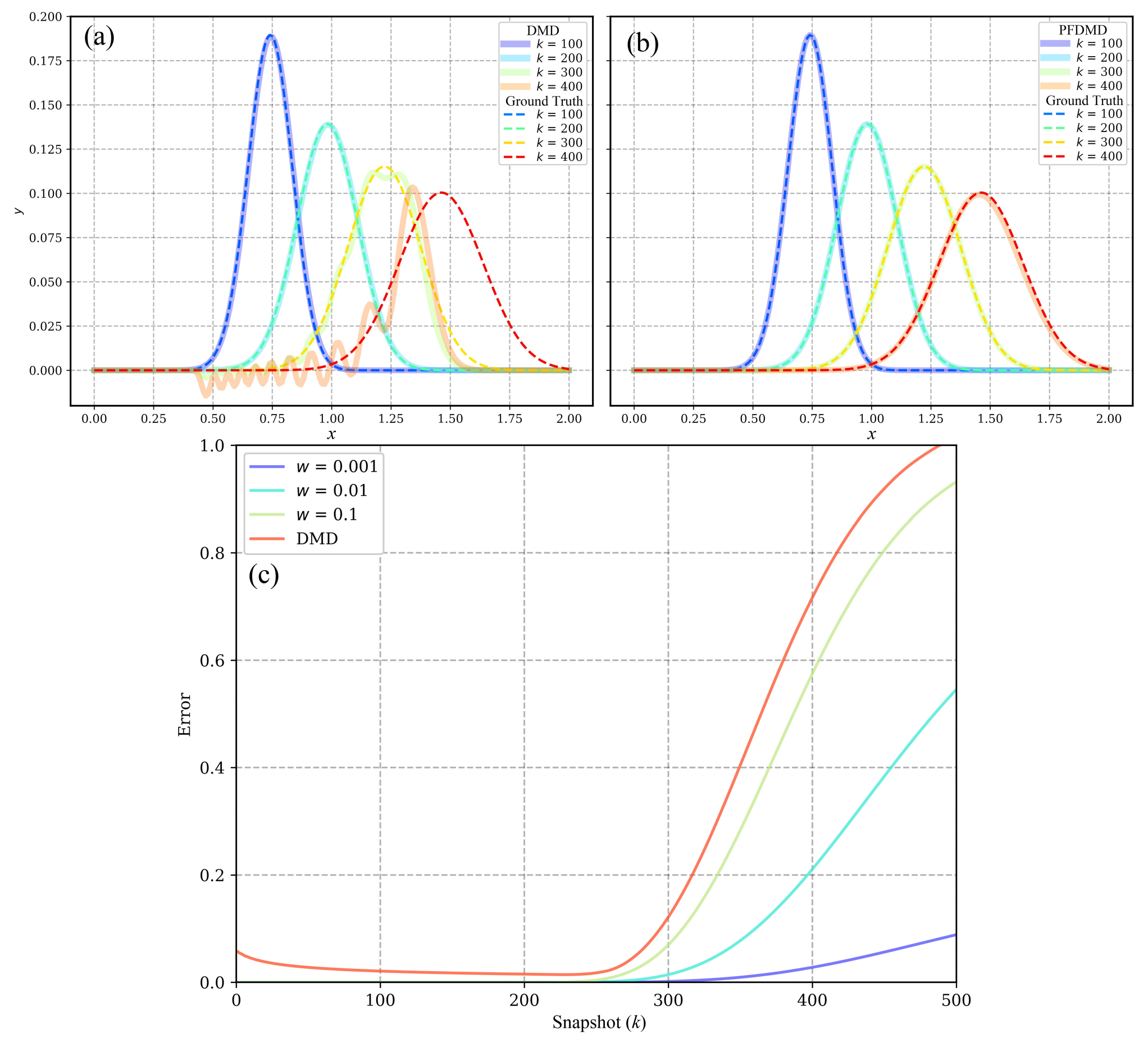}
  \caption{Advection-diffusion equation results: (a) Ground truth and prediction of DMD; (b) Ground truth and prediction of PFDMD; (c) Model error plot.}
  \label{fig:fig4}
\end{figure}

Fig. \ref{fig:fig4} demonstrates that DMD fails to capture the system dynamics, resulting in non-physical (oscillatory and negative) predictions. This failure cannot be compensated by increasing the SVD rank truncation. Therefore, it is inevitable that DMD cannot predict the translation problem beyond 200 steps, and the error increases rapidly after 200 steps. As shown in Fig.\ref{fig:fig4}(a), DMD still produces acceptable results during the reconstruction but completely fails during the prediction, with errors reaching up to 100\%. After incorporating the physical information, both the reconstruction and prediction errors of PFDMD are much lower than DMD, as shown in Fig. \ref{fig:fig4}(b) and (c). The results of reconstruction and prediction demonstrate that PFDMD effectively overcomes the limitations of DMD in predicting translation problems by incorporating physical information.

In next example, we aim to emphasize the capability of the proposed method in solving shock wave problem. Let's consider the Burgers’ equation, a nonlinear partial differential equation that plays a significant role in various fields of applied mathematics. It is commonly used to simulate the propagation of shock waves, such as in fluid mechanics, nonlinear acoustics, and gas dynamics. Here, we consider the Burgers’ equation with Dirichlet boundary conditions, specified as follows:
\begin{equation}\label{eq24}
\begin{split}
  & \frac{\partial u}{\partial t}+u\frac{\partial u}{\partial x}=\mu \frac{{{\partial }^{2}}u}{\partial {{x}^{2}}},\quad x\in [-1,1],\quad t\in [0,1] \\ 
 & b.c.\quad u\left( -1,t \right)=u\left( 1,t \right)=0, \\ 
 & i.c.\quad u\left( x,0 \right)=-\sin \left( \pi x \right), \\ 
\end{split}
\end{equation}
In this case, the spatial domain [-1,1] is discretized into \emph{m} = 200 intervals, and the time domain [0,2] is discretized into \emph{n} = 500 steps. We select \emph{n} = 150 snapshots as the training set and predict up to the $500^{th}$ step. The viscosity coefficient is set to $\mu =0.01/\pi$. The SVD truncation is 10 (rank = 10).
\begin{figure}
  \centering
  \includegraphics[width=\textwidth]{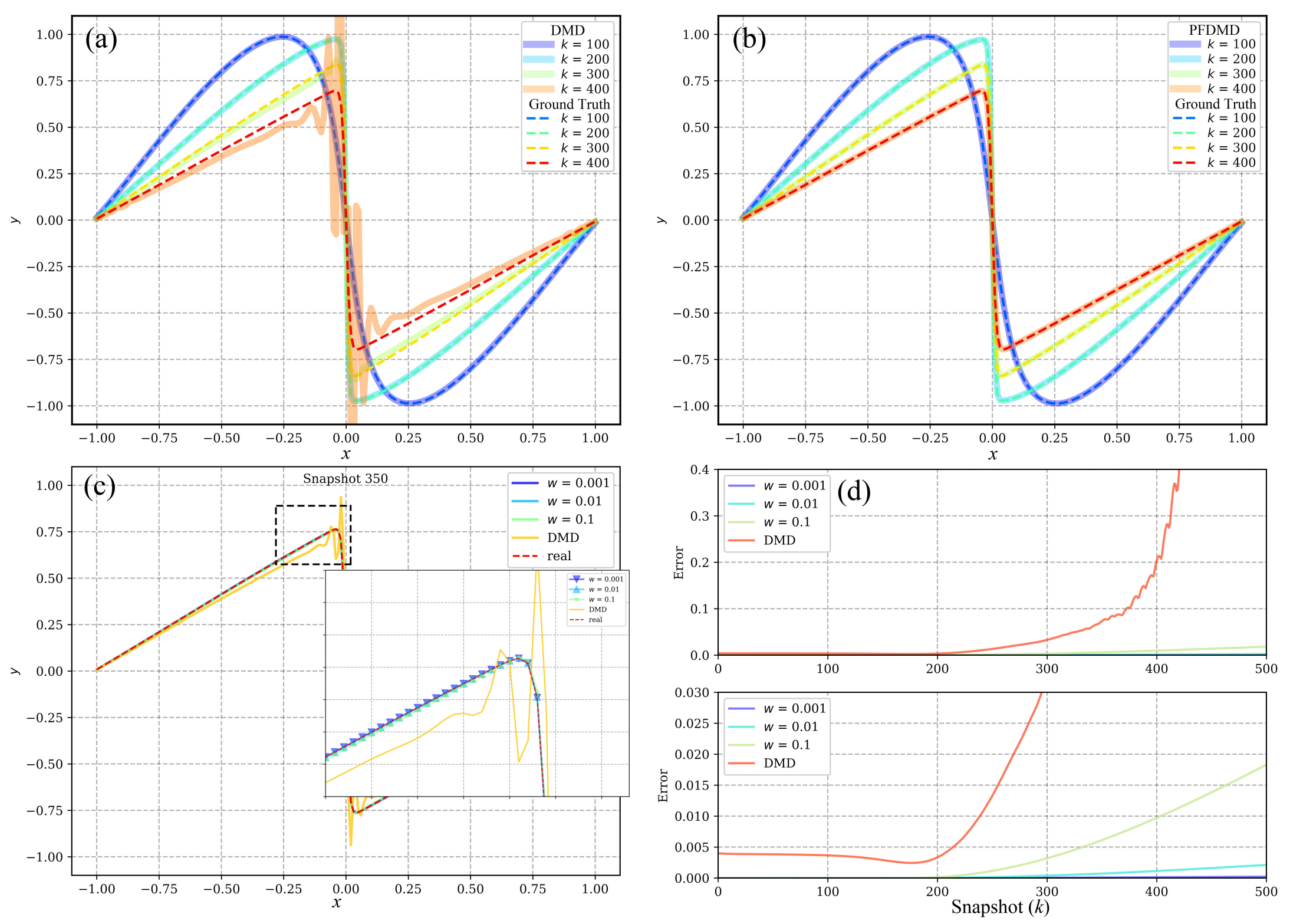}
  \caption{Burgers’ equation results: (a) Ground truth and prediction of DMD; (b) Ground truth and prediction of PFDMD; (c) Ground truth and prediction of snapshot 350; (d) Model error plot.}
  \label{fig:fig5}
\end{figure}

Fig. \ref{fig:fig5}(a) and (b) compare the reconstruction and prediction results of DMD and PFDMD. PFDMD successfully simulates the shock wave generation without non-physical oscillations at the turning points. It is worth noting that the prediction results of DMD exhibit severe oscillations after 300 steps, while PFDMD with a higher weight of physical information still produces smooth curves at the turning points. Fig. \ref{fig:fig5}(c) and (d) display the detailed $350^{th}$ prediction results and error curves. The predicted results with the chosen weights in PFDMD are more consistent with the physical situation than the DMD method, indicating that PFDMD effectively integrates physical information, keeping the prediction errors consistently lower than DMD.

Extending the one-dimensional Burgers’ equation to two-dimensional significantly increases the data amount needed for PFDMD. The following conditions are considered to investigate the applicability of this method in high-dimensional situations:
\begin{equation}\label{eq25}
\begin{split}
 & \frac{\partial u}{\partial t}+u\frac{\partial u}{\partial x}+v\frac{\partial u}{\partial y}=\mu \left( \frac{{{\partial }^{2}}u}{\partial {{x}^{2}}}+\frac{{{\partial }^{2}}u}{\partial {{y}^{2}}} \right),\quad x\in [0,2],\quad y\in [0,2], \\ 
 & \frac{\partial v}{\partial t}+u\frac{\partial v}{\partial x}+v\frac{\partial v}{\partial y}=\mu \left( \frac{{{\partial }^{2}}v}{\partial {{x}^{2}}}+\frac{{{\partial }^{2}}v}{\partial {{y}^{2}}} \right),\quad t\in [0,1] \\ 
 & b.c.\quad u\left( 0,y,t \right)=u\left( 2,y,t \right)=0,\quad u\left( x,0,t \right)=u\left( x,2,t \right)=0, \\ 
 & \quad \,\quad v\left( 0,y,t \right)=v\left( 2,y,t \right)=0,\quad v\left( x,0,t \right)=v\left( x,2,t \right)=0, \\ 
 & i.c.\quad u\left( x,y,0 \right)=v\left( x,y,0 \right)=1.2\left( \sin \left( 0.02\pi x \right)-\sin \left( 0.02\pi y \right) \right),\quad  \\ 
\end{split}
\end{equation}

\begin{figure}
  \centering
  \includegraphics[width=\textwidth]{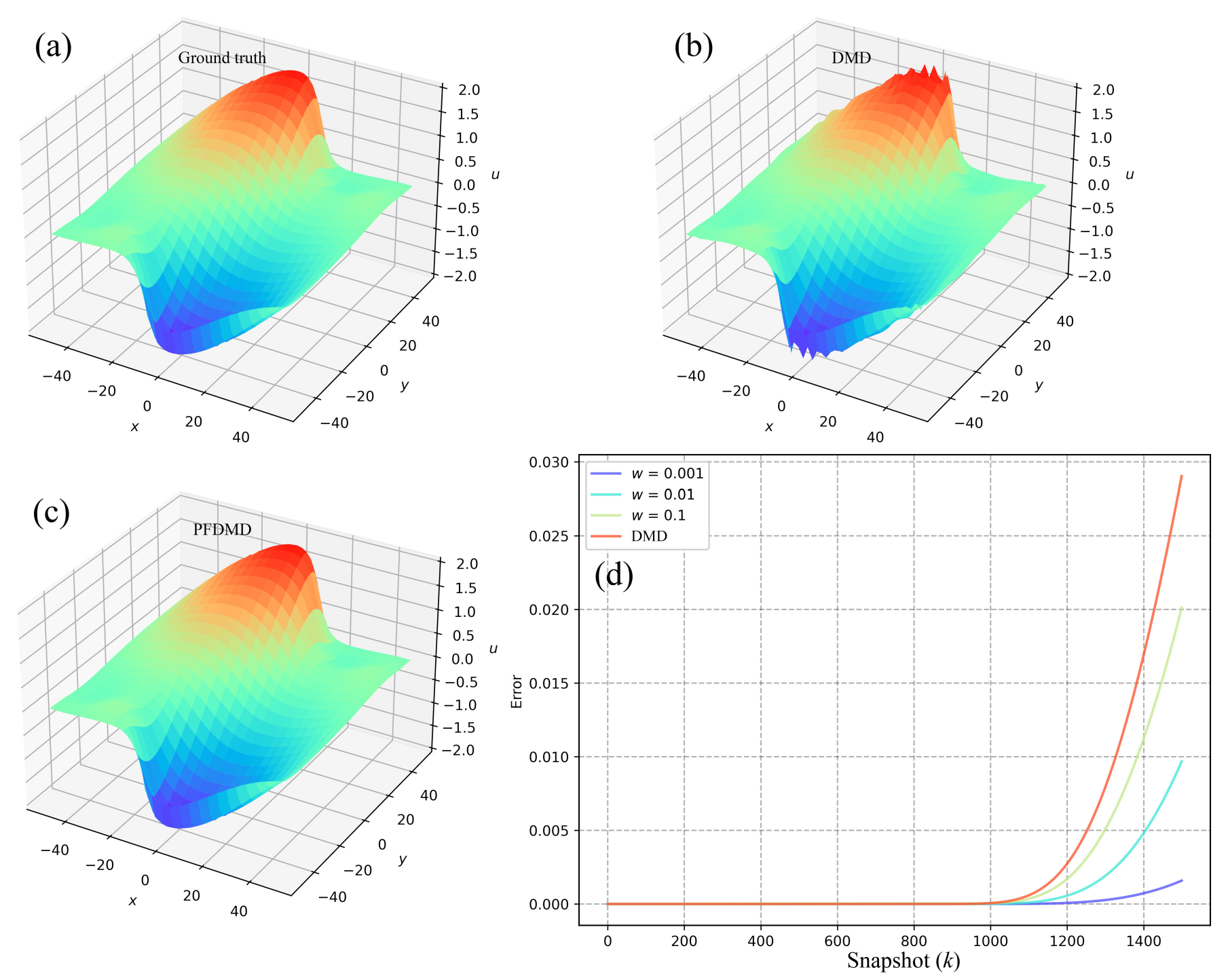}
  \caption{Two-dimensional Burgers’ equation results: (a) Ground truth of snapshot 1500; (b) DMD prediction of snapshot 1500; (c) PFDMD prediction of snapshot 1500; (d) Model error plot.}
  \label{fig:fig6}
\end{figure}

As shown in Fig. \ref{fig:fig6}(a), (b), and (c), the flow field plots of the \emph{x}-direction velocity \emph{u} in the Burgers’ equation are presented. The dataset includes the first 800 steps, and the prediction is made up to the $1500^{th}$ step. From Fig. \ref{fig:fig6}(b), it can be observed that the prediction of DMD still produces non-physical results. Not only does the predicted wave peak value end up being 0.2 higher than the actual value, but it also generates oscillations within the region where the wave peak. However, for PFDMD with the appropriate weight, the oscillation does not appear in the prediction results, and the peak value is closer to the actual value. Fig. \ref{fig:fig6}(d) shows the model error curves of different weight coefficients. It can be observed that both methods predicted have low errors to 1000 steps. However, beyond 1000 steps, the error accumulates rapidly, indicating the failure of the data-driven model at that point. Therefore, it is necessary to correct the predicted results using physical equation. Hence, lowering the weight coefficient can minimize the dependence on the data-driven model, allowing for predictions of more steps within a reasonable error range.

\subsection{Impact of inaccurate data model and physical information}

\begin{figure}
  \centering
  \includegraphics[width=\textwidth]{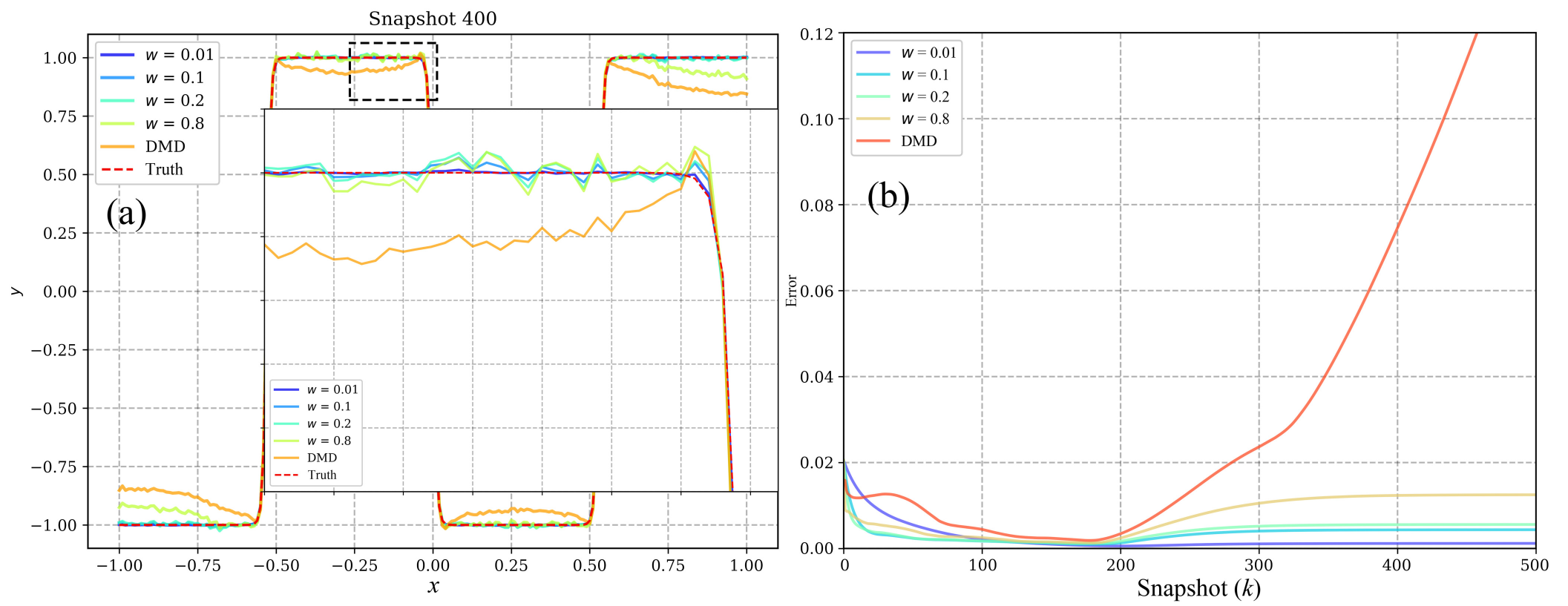}
  \caption{Allen-Cahn equation results with noise data: (a) Ground truth and prediction of snapshot 400; (b) Model error plot.}
  \label{fig:fig7}
\end{figure}

In the above example, where the physical information is completely accurate, it is evident that a higher weight of physical information leads to a smaller prediction error. However, in practical applications, data and physical information may deviate from the accurate solution. These deviations always involve noisy data, inaccurate physical information and an absence of data model. In such cases, a smaller weight is not always better. Therefore, we initially examine the impact of a noisy dataset on PFDMD by Allen-Cahn equation. In this scenario, to demonstrate the effectiveness of PFDMD, the \emph{A} matrix is generated using noisy data (original data + 0.01 * Gaussian noise). As shown in Fig. \ref{fig:fig7}(a), PFDMD can produce better prediction results and avoid the deviation and sinking state observed in DMD. This suggests that fusing physics equation can, to some extent, prevent the generation of predictions that are not consistent with physical principles. Fig. \ref{fig:fig7}(b) illustrates that due to the influence of data noise, the prediction error of DMD increases continuously and eventually exceeds 10\%. In contrast, PFDMD effectively suppresses the accumulation of prediction errors.

Next, let's consider the scenario that the data and physical equation are both inaccurate. Firstly, in one-dimensional diffusion equation, when generating the data used for DMD training, the diffusion coefficient is increased from 0.01 (the value in the accurate solution) to 0.0105, representing the deviation between the data and the accurate solution. Simultaneously, the diffusion coefficient in the physical equation is changed to 0.0095 to introduce biased physical information into PFDMD.
\begin{figure}
  \centering
  \includegraphics[width=\textwidth]{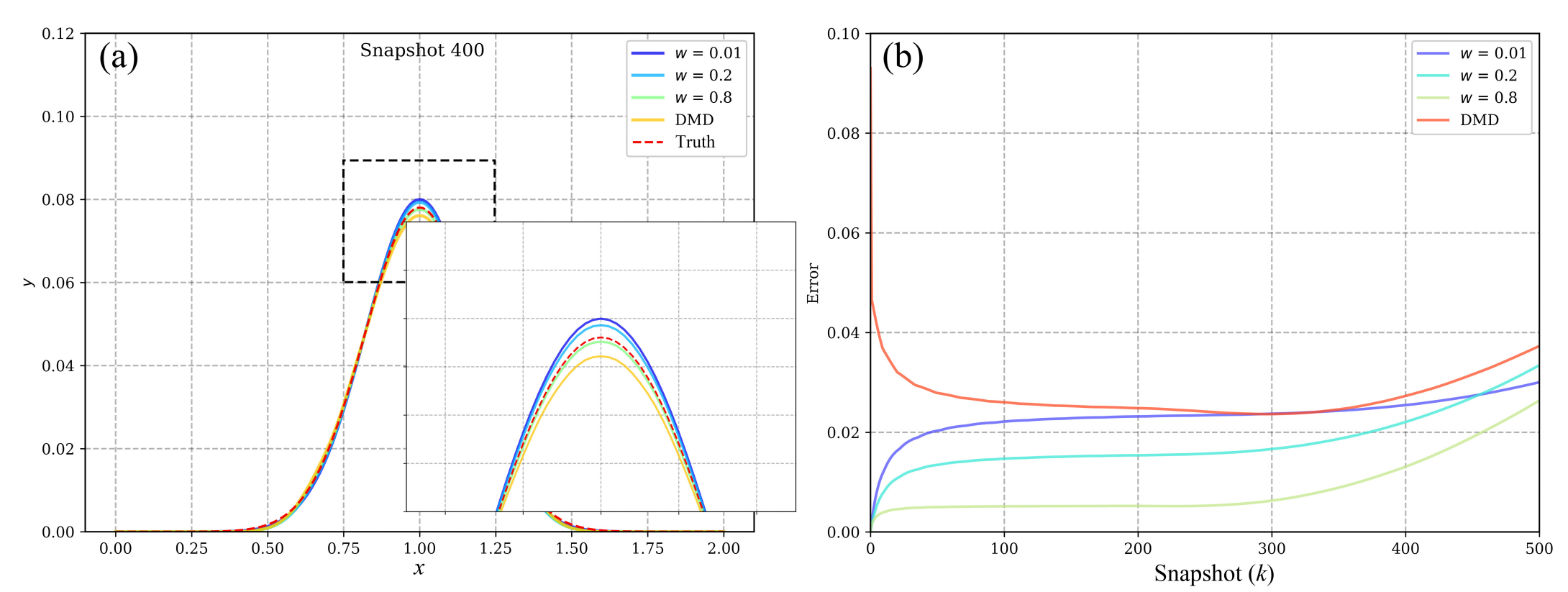}
  \caption{Diffusion equation results with inaccurate data and physical equation: (a) Ground truth and prediction of snapshot 400; (b) Model error plot.}
  \label{fig:fig8}
\end{figure}

The computational results are shown in Fig. \ref{fig:fig8}(a). The results at the peak in Fig. \ref{fig:fig8}(a) indicate that the predicted results obtained by DMD are smaller than the accurate solution due to the larger diffusion coefficient. At the same time, PFDMD demonstrates better prediction capability after incorporating the physical information. However, due to the deviation of the diffusion coefficient in physical information, the smaller the weight coefficient (i.e., the closer it is to the physical information), the larger the predicted values. The prediction error plot in Fig. \ref{fig:fig8}(b) shows that the DMD model consistently maintains a relatively high error due to the influence of inaccurate data. However, because of the inaccuracy in the physical information, when the weight is set to 0.8, the obtained results have a smaller error compared to the case of a 0.01 weight. Therefore, smaller weight does not lead to more accurate results when physical information is inaccurate. Determining weight value needs to strike a balance between data-driven and physical information-driven. 

\begin{figure}
  \centering
  \includegraphics[width=\textwidth]{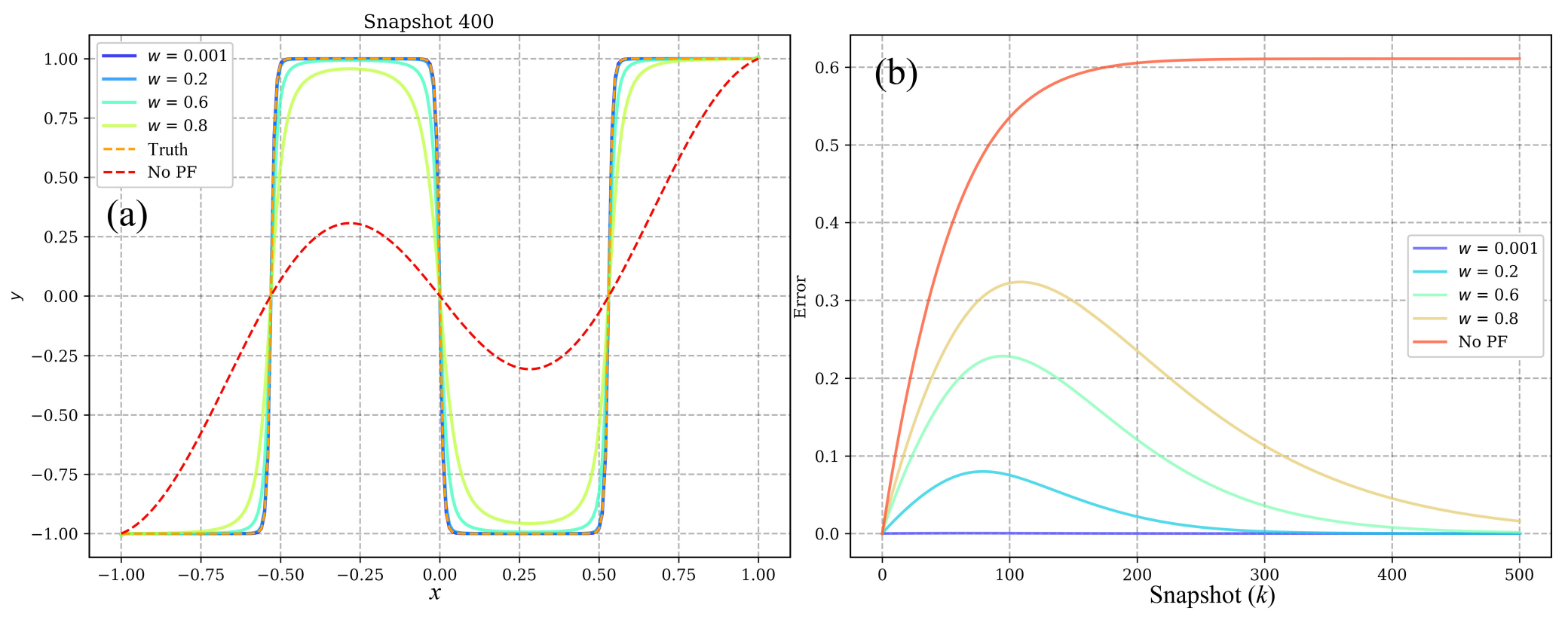}
  \caption{Allen-Cahn equation results with the case of $\emph{A} = 1$: (a) Ground truth and prediction of snapshot 400; (b) Model error plot.}
  \label{fig:fig9}
\end{figure}

Instead of using existing data to generate the \emph{A} matrix through DMD, a scenario involves where only physical information is available, without any data, by setting \emph{A} as the identity matrix. In this situation, the algorithm framework devolves into a framework driven solely by physical equations. Fig. \ref{fig:fig9}(a) shows the predicted results at the $300^th$ step of Allen-Cahn equation. When the \emph{A} matrix is the identity matrix, the result is depicted by the red curve, which represents the distribution of \emph{u} that does not change over time. However, corrections are made to the results after incorporating the physical information, regardless of the weight value \emph{w}. The smaller the weight, the smaller the influence of the identity matrix \emph{A}, and the more accurate the results, as indicated by the error curve in Fig. \ref{fig:fig9}(b). When \emph{w} is set to 0.001, very close results to the true solution can be obtained. In this case, the PFDMD actually devolves into a corrected non-iterative calculation of the PDE equation. This demonstrates that PFDMD can be applied to special cases without a data-driven method.

\subsection{Analysis of Kalman gain updating and computational cost}

\begin{figure}
  \centering
  \includegraphics[width=\textwidth]{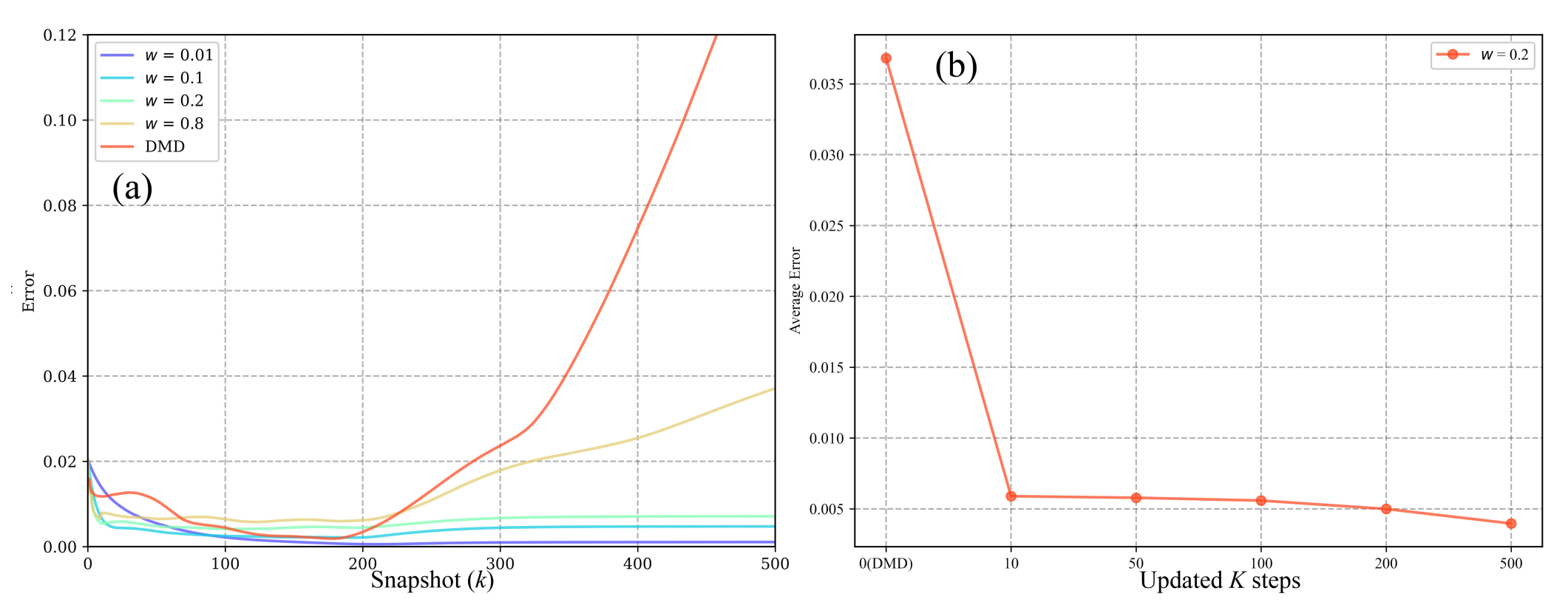}
  \caption{Allen-Cahn equation results: (a) Model error plot for the case of updating 10 timesteps \emph{K}; (b) The reconstruction and prediction average error curve of updating different timesteps \emph{K}.}
  \label{fig:fig10}
\end{figure}

One characteristic of the Kalman filtering method is that it can update the Kalman gain (\emph{K}) for only a few steps and then cease to update the \emph{K} values, thereby accelerating the computation. PFDMD can also perform such calculation. The following scenario is performed when the Kalman gain coefficient is only updated by using Eq. (\ref{eq16}) in the initial 10 timesteps. And, it is unchanged in the subsequent prediction calculations. In other words, the gain coefficients are only applied at each time step, as shown in Eq. (\ref{eq14}), without the updating step Eqs. (\ref{eq15}) to (\ref{eq18}), thereby reducing the computational complexity of PFDMD. To demonstrate the predictive capability of PFDMD after updating the Kalman gain (\emph{K}) for a limited number of steps, a dataset with significant prediction errors is required. Therefore, we continue with the noisy example of Allen-Cahn equation presented in Section 3.3. As shown in Fig. \ref{fig:fig10}(a), even after 10 correction steps, PFDMD can still produce better prediction results without updating the Kalman gain coefficients and avoid the non-physical oscillation observed in DMD. Fig. \ref{fig:fig10}(b) displays the average errors in reconstruction and prediction obtained by updating the Kalman gain (\emph{K}) over various time steps. It can be observed that continuously updating \emph{K} leads to improved prediction accuracy; however, the trade-off is an increase in computational cost.

The improved prediction results by correction are satisfactory, however, the computational cost of the algorithm should be discussed. According to the case in this section, the time taken to generate the DMD modes and the time taken by different methods to predict one step are calculated. As shown in Table ~\ref{tab:table1}, the DMD still exhibits a very low computational cost, while the updating and correction of \emph{P} and \emph{K} in PFDMD greatly increase the time of the prediction step. The prediction time of PFDMD is still controlled within the acceptable range, which is related to the small dataset size of this case. Without updating \emph{K}, the calculation time of PFDMD is close to DMD. Therefore, to reduce the calculation cost, we can stop updating K after a certain step to accelerate the prediction.
\begin{table}
 \caption{Average time of DMD modes generation and one step data-driven prediction}
  \centering
  \resizebox{\linewidth}{!}{
  \begin{tabular}{cccccc}
    \toprule
    Scenarios & DMD modes generation (rank = 10) & DMD modes generation (rank = 50) & DMD prediction & PFDMD prediction & PFDMD prediction without update \emph{K} \\
    \midrule
    Average time (s) & 0.009  & 0.019  & 0.004 & 0.077 & 0.005   \\
    \bottomrule
  \end{tabular}}
  \label{tab:table1}
\end{table}

\section{Conclusion}
This paper proposes a Physics-Fusion Dynamic Mode Decomposition (PFDMD) method, which can introduce physics information through the residual of partial differential equations. By comparing the predictive results between DMD and PFDMD in various benchmark tests, we demonstrate that PFDMD outperforms DMD noticeably with different physics information weights. The PFDMD method is capable of solving problems related to translation and shock wave. In particular, PFDMD can adjust the fusion degree of physics information to improve the versatility of the DMD model. It is worth noting that even with the combination of imperfect physics information and an inaccurate DMD model, the PFDMD can still improve prediction accuracy. 

It is worth noting that the physics fusion framework proposed in this study is not confined to DMD alone. The adaptability of this framework extends to a broader range of data-driven models, including neural networks. Fusing this framework with other data-driven models offers a promising avenue to enhance model performance.

\section*{Acknowledgments}
Financial support from the National Natural Science Foundation of China (Grant no. 22308251, 22178247, 22378304) is gratefully acknowledged.

\section*{Data availability}
The data that support the findings of this study are available from the corresponding author upon reasonable request.

\bibliographystyle{unsrt}
\bibliography{references}  

\end{document}